\documentclass{revtex4-2}%
\usepackage[latin9]{inputenc}
\usepackage{amsmath}
\usepackage{amssymb}
\usepackage{graphicx}
\usepackage[unicode=true,  bookmarks=false,  breaklinks=false,pdfborder={0 0
1},backref=section,colorlinks=false]{hyperref}
\usepackage{amsfonts}
\usepackage{subfigure}
\usepackage{cleveref}
\usepackage{listings}
\usepackage{xcolor}
\usepackage{array}
\usepackage{url}%
\hypersetup{
 colorlinks,linkcolor=blue,citecolor=blue,urlcolor=blue}
\makeatletter
\@ifundefined{textcolor}{}
{
\definecolor{BLACK}{gray}{0}
 \definecolor{WHITE}{gray}{1}
 \definecolor{RED}{rgb}{1,0,0}
 \definecolor{GREEN}{rgb}{0,1,0}
 \definecolor{BLUE}{rgb}{0,0,1}
 \definecolor{CYAN}{cmyk}{1,0,0,0}
 \definecolor{MAGENTA}{cmyk}{0,1,0,0}
 \definecolor{YELLOW}{cmyk}{0,0,1,0}
}

\makeatother
\begin{document}
\preprint{CTP-SCU/2021022}
\title{Thermodynamics and Phase Structure of an Einstein-Maxwell-scalar Model in
Extended Phase Space}
\author{Guangzhou Guo}
\email{guangzhouguo@stu.scu.edu.cn}
\author{Peng Wang}
\email{pengw@scu.edu.cn}
\author{Houwen Wu}
\email{iverwu@scu.edu.cn}
\author{Haitang Yang}
\email{hyanga@scu.edu.cn}
\affiliation{Center for Theoretical Physics, College of Physics, Sichuan University,
Chengdu, 610064, China}

\begin{abstract}
In this paper, we study thermodynamics and phase structure of asymptotically
AdS hairy and Reissner-Nordstr\"{o}m-AdS (RNAdS) black holes in the extended
phase space, where the cosmological constant is interpreted as a thermal
pressure. The RNAdS and hairy black holes are black hole solutions of an
Einstein-Maxwell-scalar (EMS) model with a non-minimal coupling between the
scalar and electromagnetic fields. The Smarr relation, the first law of
thermodynamics and the free energy are derived for black hole solutions in the
EMS model. Moreover, the phase structure of the RNAdS and hairy black holes is
investigated in canonical and grand canonical ensembles. Interestingly, RNAdS
BH/hairy BH/RNAdS BH reentrant phase transitions, consisting of zeroth-order
and second-order phase transitions, are found in both ensembles.

\end{abstract}
\maketitle
\tableofcontents

{}

\section{Introduction}

The first observations of gravitational waves by LIGO \cite{Abbott:2016blz}
and the first image of a black hole in the galaxy M87 \cite{Akiyama:2019cqa}
have ushered us into a new era of black hole physics. Black hole
thermodynamics has been a hot topic for research in black hole physics since
the pioneering work \cite{Hawking:1974sw,Bekenstein:1972tm,Bekenstein:1973ur},
where Hawking and Bekenstein found that black holes can possess temperature
and entropy. Analogous to the laws of thermodynamics, the four laws of black
hole mechanics were established in \cite{Bardeen:1973gs}.

Asymptotically AdS black holes can be in thermal equilibrium with the thermal
radiation since the AdS boundary serves as a reflecting wall for the thermal
radiation. Therefore, it is proper to study black hole thermodynamics for AdS
black holes. Indeed, thermodynamic properties of AdS black holes were first
investigated in \cite{Hawking:1982dh}, where the Hawking-Page phase transition
between Schwarzschild AdS black holes and the thermal AdS space was
discovered. With the advent of the AdS/CFT correspondence
\cite{Maldacena:1997re,Gubser:1998bc,Witten:1998qj}, there has been much
interest in studying thermodynamics and phase structure of AdS black holes
\cite{Witten:1998zw,Cvetic:1999ne,Chamblin:1999tk,Chamblin:1999hg,Caldarelli:1999xj,Cai:2001dz,Cvetic:2001bk,Nojiri:2001aj,Astefanesei:2019ehu}%
. In particular, RNAdS black holes were observed to possess a van der
Waals-like phase transition (i.e., a phase transition consisting of a
first-order phase transition terminating at a second-order critical point) in
a canonical ensemble \cite{Chamblin:1999tk,Chamblin:1999hg} and a
Hawking-Page-like phase transition in a grand canonical ensemble
\cite{Peca:1998cs}. Alternatively, asymptotically flat black holes can be
thermally stable with a dilaton potential \cite{Astefanesei:2019mds,Astefanesei:2020xvn}, or by placing them inside a spherical cavity, on the wall of
which the metric is fixed \cite{York:1986it}. Early studies often concluded
that black holes in a cavity have quite similar phase structure and
transitions to the AdS counterparts
\cite{Braden:1990hw,Carlip:2003ne,Lundgren:2006kt,Wang:2019urm}. However, it
has recently been found that black holes in a cavity and the AdS counterparts
can show different thermodynamic behavior, e.g., phase structure of
Born-Infeld black holes \cite{Wang:2019kxp,Liang:2019dni}, thermodynamic
geometry of RN black holes \cite{Wang:2019cax} and the second law of
thermodynamics \cite{Wang:2020osg}.

However, unlike everyday thermodynamics, there is no pressure or volume
associated with a black hole in the aforementioned black hole thermodynamics.
To address this issue, the cosmological constant can be treated as a
thermodynamic variable and interpreted as a thermodynamic pressure, which
leads to the thermodynamics of black holes in the extended phase space
\cite{Dolan:2011xt,Kubiznak:2012wp,Kubiznak:2016qmn}. The mass of black holes
is then identified as the chemical enthalpy rather than the internal energy
due to a pressure-volume term $V\delta P$ in the first law of thermodynamics
\cite{Kastor:2009wy}. The thermodynamic behavior has been investigated for
various AdS black holes in the extended phase space
\cite{Wei:2012ui,Gunasekaran:2012dq,Cai:2013qga,Xu:2014kwa,Frassino:2014pha,Dehghani:2014caa,Hennigar:2015esa,Caceres:2015vsa,Wei:2015iwa,Hendi:2016yof,Hendi:2017fxp,Lemos:2018cfd,Pedraza:2018eey,Wang:2018xdz,Wei:2019uqg,Wei:2020poh}%
, which discovered a broad range of new phenomena. For a recent review, see
\cite{Kubiznak:2016qmn}. Especially for a RNAdS black hole, the coexistence
line of the Hawking-Page phase transition in the $P$-$T$ diagram is
semi-infinite and reminiscent of the solid/liquid phase transition in a grand
canonical ensemble \cite{Kubiznak:2014zwa}, while the coexistence line in the
$P$-$T$ diagram is finite and terminates at a critical point in a canonical
ensemble \cite{Kubiznak:2012wp}. Intriguingly, more novel phase behavior has
been found for more complicated black hole spacetimes. For example, reentrant
phase transitions, which are composed of zeroth-order and first-order phase
transitions, were discovered in Born-Infeld-AdS black holes
\cite{Gunasekaran:2012dq,Wang:2018xdz}, higher dimensional singly spinning
Kerr-AdS black holes \cite{Altamirano:2013ane}, AdS black holes in Lovelock
gravity \cite{Frassino:2014pha}, AdS black holes in dRGT massive gravity
\cite{Zou:2016sab}, and hairy AdS black holes \cite{Hennigar:2015wxa}. A
reentrant phase transition depicts a phenomenon that a system undergoes at
least two phase transitions and returns to the macroscopically similar initial
state under the monotonic variation of a thermodynamic variable. This
phenomenon was first observed in a nicotine/water mixture, where a homogeneous
mixture state transforms to a distinct nicotine/water phase as the temperature
increases, and eventually return to the mixture state at a sufficiently high
temperature \cite{hudson1904gegenseitige}. Moreover, since the primary
motivation to study AdS black holes is the AdS/CFT correspondence, there have
been some interesting works understanding the extended phase space
thermodynamics in the framework of the AdS/CFT correspondence
\cite{Karch:2015rpa,AlBalushi:2020rqe,Cong:2021fnf}. It is noteworthy that the
thermodynamics and phase structure of black holes in a cavity have been
investigated in the extended phase space \cite{Simovic:2018tdy,Wang:2020hjw}.

The no-hair theorem states that a black hole can be uniquely determined via
its mass, electric charge and angular momentum
\cite{Israel:1967wq,Carter:1971zc,Ruffini:1971bza}. Although this theorem can
be proven when subjected to some specific energy conditions, e.g., the
Einstein-Maxwell theory, various hairy black holes have been constructed to
provide counter-examples to the no-hair theorem
\cite{Volkov:1989fi,Bizon:1990sr,Greene:1992fw,Luckock:1986tr,Droz:1991cx,Kanti:1995vq,Mahapatra:2020wym}%
. For a review, see \cite{Herdeiro:2015waa}. Since testing the no-hair theorem
is crucial to understand black hole physics, it is of great interest to find
hairy black hole solutions and study their properties. To understand the
formation of hairy black holes, authors of \cite{Herdeiro:2018wub} proposed an
Einstein-Maxwell-scalar (EMS) model with a non-minimal coupling between the
scalar and electromagnetic fields, and studied the phenomenon of spontaneous
scalarization in the model. Subsequently, various properties of this model and
its extensions were discussed in the literature, e.g., different non-minimal
coupling functions \cite{Fernandes:2019rez,Blazquez-Salcedo:2020nhs}, dyons
including magnetic charges \cite{Astefanesei:2019pfq}, axionic-type couplings
\cite{Fernandes:2019kmh}, massive and self-interacting scalar fields
\cite{Zou:2019bpt,Fernandes:2020gay}, horizonless reflecting stars
\cite{Peng:2019cmm}, stability analysis of hairy black holes
\cite{Myung:2018vug,Myung:2019oua,Zou:2020zxq,Myung:2020etf,Mai:2020sac},
higher dimensional scalar-tensor models \cite{Astefanesei:2020qxk},
quasinormal modes of hairy black holes
\cite{Myung:2018jvi,Blazquez-Salcedo:2020jee}, two U(1) fields
\cite{Myung:2020dqt}, quasi-topological electromagnetism \cite{Myung:2020ctt},
topology and spacetime structure influences \cite{Guo:2020zqm}, the
Einstein-Born-Infeld-scalar theory \cite{Wang:2020ohb}, with a negative
cosmological constant \cite{Zhang:2021etr} and images of hairy black holes
with accretion flows \cite{Gan:2021pwu,Gan:2021xdl}. In our recent work
\cite{Guo:2021zed}, we obtained asymptotically AdS hairy black hole solutions
in the EMS model and investigated the phase structure of a canonical ensemble
of RNAdS and hairy black holes in a normal phase space, where the cosmological
constant is fixed. It showed that hairy and RNAdS black holes coexist in some
parameter region, leading to a reentrant phase transition.

In this paper, focusing on the extended phase space, we discuss thermodynamic
properties and phase structure of asymptotically AdS black hole solutions in
the EMS model, which were obtained in \cite{Guo:2021zed}. The remainder of
this paper is organized as follows. In Section
\ref{sec:Thermodynamic quantities of hairy black hole}, after we briefly
review asymptotically AdS hairy black hole solutions in the EMS model, the
Smarr relation, the first law of thermodynamics and the free energy are
derived. In Section \ref{sec:phase structure and transition}, we study phase
structure and transitions of hairy and RNAdS black holes in canonical and
grand canonical ensembles by minimizing the free energy. We finally conclude
our results in Section \ref{sec:Discussion-and-Conclusions}. We take $G=1$ for
simplicity throughout this paper.

\section{Thermodynamics}

\label{sec:Thermodynamic quantities of hairy black hole}

In this section, we first briefly review asymptotically AdS hairy black hole
solutions in the EMS model. After the first law of thermodynamics is obtained
by a covariant approach, we derive the Smarr relation using dimensional
analysis and the Komar integral with respect to a time-like Killing vector.
Finally, the free energy is computed in canonical and grand canonical
ensembles, respectively, to study the phase structure and transitions of the
hairy black holes.

\subsection{Hairy Black Hole Solutions}

We consider an EMS model with a scalar field minimally coupled to the metric
field and non-minimally coupled to the electromagnetic field, which is
described by the action,
\begin{equation}
S_{\text{bulk}}=-\frac{1}{16\pi}\int d^{4}x\sqrt{-g}\left[  R-2\Lambda
-2\left(  \partial\phi\right)  ^{2}-f\left(  \phi\right)  F_{\mu\nu}F^{\mu\nu
}\right]  . \label{eq:action}%
\end{equation}
Here, $f\left(  \phi\right)  $ is a non-minimal coupling function between the
scalar field $\phi$ and the electromagnetic field $A_{\mu}$, $F_{\mu\nu
}=\partial_{\mu}A_{\nu}-\partial_{\nu}A_{\mu}$ is the electromagnetic field
strength tensor, and $\Lambda=-3/L^{2}$ is the cosmological constant with the
AdS radius $L$. As in \cite{Guo:2021zed}, we focus on the coupling function
$f\left(  \phi\right)  =e^{\alpha\phi^{2}}$ with $\alpha\geq0$ and the
spherically symmetric ansatz for the metric, the electromagnetic field and the
scalar field,
\begin{align}
ds^{2}  &  =-N\left(  r\right)  e^{-2\delta\left(  r\right)  }dt^{2}+\frac
{1}{N\left(  r\right)  }dr^{2}+r^{2}\left(  d\theta^{2}+\sin^{2}\theta
d\varphi^{2}\right)  ,\nonumber\\
A_{\mu}dx^{\mu}  &  =V\left(  r\right)  dt\text{ and }\phi=\phi\left(
r\right)  . \label{eq:metric}%
\end{align}
The equations of motion are%
\begin{align}
N^{\prime}\left(  r\right)   &  =\frac{1-N\left(  r\right)  }{r}-\frac{Q^{2}%
}{r^{3}e^{\alpha\phi^{2}\left(  r\right)  }}-rN\left(  r\right)  \phi
^{\prime2}\left(  r\right)  +\frac{3r}{L^{2}},\nonumber\\
\left(  r^{2}N\left(  r\right)  \phi^{\prime}\left(  r\right)  \right)
^{\prime}  &  =-\frac{\alpha\phi\left(  r\right)  Q^{2}}{e^{\alpha\phi
^{2}\left(  r\right)  }r^{2}}-r^{3}N\left(  r\right)  \phi^{\prime3}\left(
r\right)  ,\nonumber\\
\delta^{\prime}\left(  r\right)   &  =-r\phi^{\prime2}\left(  r\right)
,\label{eq:NLEqs}\\
V^{\prime}\left(  r\right)   &  =\frac{Q}{r^{2}e^{\alpha\phi^{2}\left(
r\right)  }}e^{-\delta\left(  r\right)  },\nonumber
\end{align}
where primes denote the derivatives with respect to the radial coordinate $r$,
and the integration constant $Q$ is interpreted as the electric charge of the
black hole solution. For later use, we introduce the Misner-Sharp mass
function $m\left(  r\right)  $ by $N\left(  r\right)  =1-2m\left(  r\right)
/r+r^{2}/L^{2}$.

To solve the set of non-linear ordinary differential equations $\left(
\ref{eq:NLEqs}\right)  $, we impose appropriate boundary conditions at the
event horizon of radius $r_{+}$ and the spatial infinity as%
\begin{align}
m(r_{+})  &  =\frac{r_{+}}{2}+\frac{r_{+}^{3}}{2L^{2}}\text{, }\delta
(r_{+})=\delta_{0}\text{, }\phi(r_{+})=\phi_{0}\text{, }V(r_{+})=0\text{,}%
\nonumber\\
m(\infty)  &  =M\text{, }\delta(\infty)=0\text{, }\phi(\infty)=0\text{,
}V(\infty)=\Phi\text{,} \label{eq:BC}%
\end{align}
where $\delta_{0}$ and $\phi_{0}$ are two positive constants, $M$ is the ADM
mass, and $\Phi$ is the electrostatic potential. Note that the general
asymptotic scalar field solution is $\phi\left(  r\right)  \sim\phi_{-}%
+\frac{\phi_{+}}{r^{3}}$, where $\phi_{+}$ can be interpreted as the
expectation value of the dual operator of the scalar field on the conformal
boundary in the presence of the external source $\phi_{-}$. In this paper, we
assume the absence of the external source, i.e., $\phi_{-}=0$, which has been
used in holographic applications with spontaneous symmetry breaking, e.g.,
holographic superconductors \cite{Hartnoll:2008vx} and holographic superfluids
\cite{Herzog:2008he}. Usually, the shooting method is used to obtain hairy
black hole solutions of the non-linear differential equations $\left(
\ref{eq:NLEqs}\right)  $, which satisfy the boundary conditions $\left(
\ref{eq:BC}\right)  $. Here, we use the NDSolve function in Wolfram
Mathematica to numerically solve the differential equations $\left(
\ref{eq:NLEqs}\right)  $.

The hairy black hole solutions can be characterized by the number $n$ of nodes
of the scalar field. In \cite{Guo:2021zed}, it was demonstrated that the
fundamental branch of the solutions, i.e., $n=0$, is stable against radial
perturbations, whereas the $n=1$ and $2$ excited branches are not. Therefore,
we focus on the fundamental branch in this paper. Note that RNAdS black holes
also are solutions of the EMS model $\left(  \ref{eq:action}\right)  $. With
increasing the charge-to-mass ratio $q\equiv Q/M$ , the RNAdS black hole
solution becomes unstable against the scalar field perturbation
\cite{Guo:2021zed}. At the onset of the instability (dubbed the bifurcation
line), there appears a zero mode of the scalar field perturbation, which can
be extended to a nonlinear regime and gives black hole solutions with a scalar
hair. In addition, the hairy black holes coexist with RNAdS black holes in
some region of parameter space.

\subsection{First Law of Thermodynamics and Thermodynamic Volume}

In the extended phase space, we use the Iyer and Wald's covariant construction
\cite{Lee:1990nz,Wald:1993nt,Iyer:1994ys,Urano:2009xn,Wu:2016auq} to obtain
the first law of thermodynamics for hairy black holes. We start with a 4-form
Lagrangian $\mathbf{L}$, which is diffeomorphism invariant and satisfies
\begin{equation}
\mathbf{L}\left(  f^{\ast}\phi\right)  =f^{\ast}\left(  \mathbf{L}\left(
\phi\right)  \right)  .\label{eq:diffeomorphism}%
\end{equation}
Here, $\phi$ collectively denotes various fields, including the metric
$g_{\mu\nu}$, the electromagnetic field $A_{\mu}$ and other dynamical fields,
$f^{\ast}$ represents the pullback after a diffeomorphism map $f$, and $\ast$
is the Hodge star operator. Alternatively, an equivalent description of
diffeomorphism invariant $\left(  \ref{eq:diffeomorphism}\right)  $ is
\begin{equation}
\delta_{\xi}\mathbf{L}=\mathcal{L}_{\xi}\mathbf{L}=\ast E\mathcal{L}_{\xi}%
\phi+d\left(  \ast\theta\right)  ,\label{eq:infinitesimal diffeomorphism}%
\end{equation}
which relates the variation alone a vector field $\xi$ to a corresponding Lie
derivative $\mathcal{L}_{\xi}$. Here, $E$ schematically denotes the equations
of motion with respect to $\phi$, and the symplectic potential form $\theta$
is an one-form. Subsequently, we define a current
\begin{equation}
\ast j_{\xi}=\ast\theta\left(  \phi,\mathcal{L}_{\xi}\phi\right)  -\xi
\cdot\mathbf{L}.\label{eq:Noether current}%
\end{equation}
By virtue of eqn. $\left(  \ref{eq:infinitesimal diffeomorphism}\right)  $, an
exterior derivative acting on the current $\left(  \ref{eq:Noether current}%
\right)  $ then yields
\begin{equation}
d(\ast j_{\xi})=-\ast E\mathcal{L}_{\xi}\phi,\label{eq:conserved current}%
\end{equation}
which shows that the current is conserved if the equations of motion are
satisfied. In particular, this current is thus referred to as the Noether
current associated with the diffeomorphism symmetry. Consequently, there is a
2-form Noether charge $\ast Q_{\xi}$ related to the vector field $\xi$, which
is constructed by
\begin{equation}
\ast j_{\xi}=d\left(  \ast Q_{\xi}\right)  .\label{Noether charge}%
\end{equation}
In general, a symplectic form $\omega\left(  \phi,\delta_{1}\phi,\delta
_{2}\phi\right)  $ can be built up with an one-form $\theta\left(  \phi
,\delta\phi\right)  $,
\begin{equation}
\omega\left(  \phi,\delta_{1}\phi,\delta_{2}\phi\right)  \equiv\delta
_{2}\left(  \ast\theta\left(  \phi,\delta_{1}\phi\right)  \right)  -\delta
_{1}\left(  \ast\theta\left(  \phi,\delta_{2}\phi\right)  \right)
.\label{eq:ssf_omega}%
\end{equation}
Replacing one of variations with the Lie derivative, i.e., $\delta
\rightarrow\mathcal{L}_{\xi}$, gives the special symplectic form
$\omega\left(  \phi,\delta\phi,\mathcal{L}_{\xi}\phi\right)  $,
\begin{equation}
\omega\left(  \phi,\delta\phi,\mathcal{L}_{\xi}\phi\right)  =d\left(
\delta\left(  \ast Q\right)  -i_{\xi}\left(  \ast\theta\left(  \phi,\delta
\phi\right)  \right)  \right)  +i_{\xi}\left(  \ast E\delta\phi\right)
.\label{eq:special symplectic form}%
\end{equation}
Furthermore, integrating this special symplectic form over a Cauchy surface
$\Sigma$ leads to the variation of Hamiltonian \cite{Iyer:1994ys,Wu:2016auq},
\begin{equation}
\delta H_{\xi}=\int_{\Sigma}\omega\left(  \phi,\delta\phi,\mathcal{L}_{\xi
}\phi\right)  =\int_{\partial\Sigma}\left(  \delta\left(  \ast Q_{\xi}\right)
-\xi\cdot\left(  \ast\theta\right)  \right)  +\int_{\Sigma}\xi\cdot\left(
\ast E\delta\phi\right)  .\label{eq:variation of Hamiltonian}%
\end{equation}
In practice, the Cauchy surface $\Sigma$ is generally chosen as a constant
time hypersurface, whose boundary $\partial\Sigma$ is composed of the event
horizon and the spatial infinity. The variation of Hamiltonian $\delta H_{\xi
}$ should vanish if $\xi$ is a Killing vector, i.e., $\mathcal{L}_{\xi}\phi
=0$. Note that if some background variables, e.g., the cosmological constant,
are treated as dynamical fields, the associated equations of motion are not
necessarily satisfied, and hence the last term in eqn. $\left(
\ref{eq:variation of Hamiltonian}\right)  $ can make non-vanishing
contributions \cite{Urano:2009xn,Wu:2016auq}.

To derive the first law of thermodynamics for hairy black holes, we apply the
identity $\left(  \ref{eq:variation of Hamiltonian}\right)  $ to hairy black
hole solutions with a time-like Killing vector $\xi=\partial_{t}$. From eqn.
$\left(  \ref{eq:infinitesimal diffeomorphism}\right)  $, one can obtain the
expression of the symplectic potential form,
\begin{equation}
\theta^{\mu}=-\frac{1}{16\pi}\left[  \nabla^{\mu}\left(  g_{\rho\sigma}\delta
g^{\rho\sigma}\right)  -\nabla_{\nu}\left(  \delta g^{\mu\nu}\right)
-4g^{\mu\nu}\nabla_{\nu}\phi\delta\phi-4f\left(  \phi\right)  F^{\mu\nu}\delta
A_{\nu}\right]  .\label{eq:spf}%
\end{equation}
The Noether charge $Q_{\xi}$ can be deduced by the definition $\left(
\ref{Noether charge}\right)  $,
\begin{equation}
Q^{\mu\nu}=\frac{1}{16\pi}\left(  \nabla^{\mu}\xi^{\nu}-\nabla^{\nu}\xi^{\mu
}+4f\left(  \phi\right)  F^{\mu\nu}A_{\alpha}\xi^{\alpha}\right)  .\label{NC}%
\end{equation}
Substituting eqns. $\left(  \ref{eq:spf}\right)  $ and $\left(  \ref{NC}%
\right)  $ into the identity $\left(  \ref{eq:variation of Hamiltonian}%
\right)  $, one has
\begin{align}
\int_{r=\infty}\left(  \delta\left(  \ast Q_{\xi}\right)  -\xi\cdot\left(
\ast\theta\right)  \right)   &  =\left(  \delta QA_{t}+\frac{1}{2}re^{-\delta
}\delta N+r^{2}e^{-\delta}N\phi^{\prime}\delta\phi\right)  |_{r=+\infty
},\nonumber\\
\int_{r=r_{+}}\left(  \delta\left(  \ast Q_{\xi}\right)  -\xi\cdot\left(
\ast\theta\right)  \right)   &  =\int_{r=r_{+}}\delta\left(  \ast Q_{\xi
}\right)  |_{\xi=0},\\
\int_{\Sigma}\xi\cdot\left(  \ast E\delta\phi\right)   &  =\frac{1}{2}%
\int_{r_{+}}^{+\infty}drr^{2}e^{-\delta}\delta\Lambda,\nonumber
\end{align}
where we allow the cosmological constant to be varied as a dynamical field.
Since $\delta H_{\xi}=0$ for the Killing vector $\xi=\partial_{t}$, one
arrives at the first law of thermodynamics for hairy black holes in the
extended phase space,
\begin{equation}
\delta M=T\delta S+\Phi\delta Q+\left(  e^{-\delta_{0}}\frac{4\pi r_{+}^{3}%
}{3}-\int_{r_{+}}^{\infty}dr\delta^{\prime}e^{-\delta}\frac{4\pi r^{3}}%
{3}\right)  \delta P,\label{eq:FLT}%
\end{equation}
where $T\equiv e^{-\delta(r_{+})}N^{\prime}(r_{+})/4\pi$ is the Hawking
temperature, $S\equiv\pi r_{+}^{2}$ is the black hole entropy, and
$P\equiv-\Lambda/8\pi=3/8\pi L^{2}$ is the thermodynamic pressure. Note that
the mass $M$ plays the role of an enthalpy in the extended phase space.
Accordingly, the conjugate thermodynamic volume is given by
\begin{equation}
V=\left(  \frac{\partial M}{\partial P}\right)  _{S,Q}=e^{-\delta_{0}}%
\frac{4\pi r_{+}^{3}}{3}-\int_{r_{+}}^{\infty}dr\delta^{\prime}e^{-\delta
}\frac{4\pi r^{3}}{3}.\label{eq:volume}%
\end{equation}
For a RNAdS black hole with $\delta\left(  r\right)  =0$, the thermodynamic
volume reduces to $V=4\pi r_{+}^{3}/3$.

\begin{figure}[ptb]
\begin{centering}
\includegraphics[scale=1]{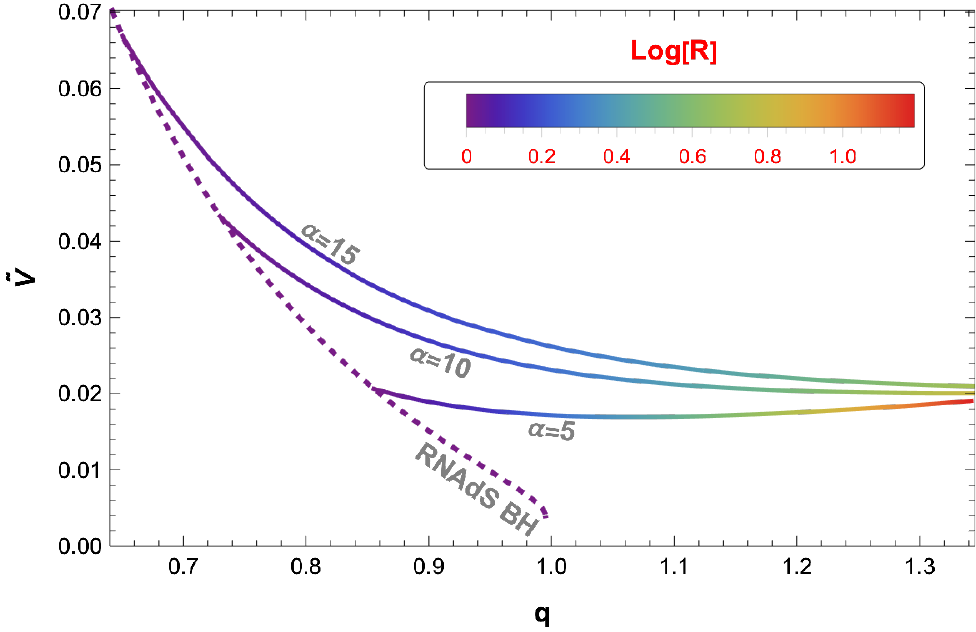}
\par\end{centering}
\caption{Reduced thermodynamic volume $\tilde{V}\equiv V/L^{3}$ versus
$q\equiv Q/M$ for RNAdS black holes and hairy black holes with $\alpha=5$,
$10$ and $15$. The color of lines specifies the value of $\log R$, where $R$
is the isoperimetric ratio. The hairy black holes have a larger thermodynamic
volume than the RNAdS black hole of the same $q$, and satisfy the reverse
isoperimetric inequality $R\geq1$.}%
\label{fig:V}%
\end{figure}

In FIG. \ref{fig:V}, we display the thermodynamic volume as a function of
$q\equiv Q/M$ for RNAdS black holes and hairy black holes with $\alpha=5$,
$10$ and $15$. It shows that hairy hole solutions bifurcate from RNAdS black
hole solutions. In the coexisting region, for a given $q$, the hairy black
holes have a larger thermodynamic volume than the RNAdS black hole. Moreover,
the thermodynamic volume of hairy black holes becomes larger for a stronger
$\alpha$. We also calculate the isoperimetric ratio,
\begin{equation}
R\equiv\left(  \frac{3V}{4\pi r_{+}^{3}}\right)  ^{\frac{1}{3}},
\end{equation}
for the RNAdS and hairy black holes, which is represented by the color of
lines in FIG. \ref{fig:V}. In fact, it was conjectured in \cite{Cvetic:2010jb}
that a reverse isoperimetric inequality $R\geq1$ holds for AdS black holes.
Interestingly, several black hole solutions have been found to violate the
reverse isoperimetric inequality
\cite{Hennigar:2014cfa,Klemm:2014rda,Hennigar:2015cja,Jing:2020sdf}. FIG.
\ref{fig:V} shows that the RNAdS and hairy black holes satisfy the reverse
isoperimetric inequality, i.e., $\log R\geq0$.

\subsection{Smarr Relation}

The Smarr relation, which relates various quantities of black holes, plays an
important role in black hole thermodynamics \cite{Smarr:1972kt}. Using the
Euler's formula for homogeneous functions, the Smarr relation can be derived
from the first law of thermodynamics $\left(  \ref{eq:FLT}\right)  $, which
relates various differential quantities. In fact, due to the Euler's theorem,
we can write the black hole mass $M=M\left(  S,P,Q\right)  $ as%
\begin{equation}
M=2\frac{\partial M}{\partial S}S-2\frac{\partial M}{\partial P}%
P+\frac{\partial M}{\partial Q}Q=2TS-2PV+\Phi Q, \label{eq:SR}%
\end{equation}
where we use $\left[  M\right]  =\left[  Q\right]  =\left[  L\right]  $,
$\left[  S\right]  =\left[  L\right]  ^{2}$ and $\left[  P\right]  =\left[
L\right]  ^{-2}$, and the partial derivatives can be expressed in terms of
black hole quantities via the first law of thermodynamics. Note that there is
no contribution from the coupling parameter $\alpha$ to the Smarr relation
since $\alpha$ is dimensionless.

Alternatively, the Smarr relation can also be derived by geometric means
\cite{Bardeen:1973gs,Kastor:2009wy}. To obtain the Smarr relation, we consider
a hypersurface $\Sigma$ with the boundary $\partial\Sigma$ in a manifold
$\mathcal{M}$ endowed with a time-like Killing vector $\xi=\partial_{t}$. Due
to Gauss's law and Einstein's equations, an identity for a Killing vector
$\nabla_{\mu}\nabla_{\nu}\xi^{\mu}=\xi^{\mu}R_{\mu\nu}$ can be integrated on
$\Sigma$,
\begin{equation}
\int_{\partial\Sigma}dS_{\mu\nu}\nabla^{\mu}\xi^{\nu}=\int_{\Sigma}dS_{\mu}%
\xi_{\nu}\left(  2T_{\mu\nu}-Tg_{\mu\nu}-\frac{3g_{\mu\nu}}{L^{2}}\right)  ,
\label{eq:Guass's law}%
\end{equation}
where $dS_{\mu\nu}$ is the surface element normal to $\partial\Sigma$ ,
$dS_{\mu}$ is correspondingly the volume element on $\Sigma$, and $T_{\mu\nu}$
is the energy-momentum tensor. For the hairy black hole, we choose the
hypersurface of constant time $t$ bounded by the event horizon and the spatial
infinity, such that the boundary $\partial\Sigma$ consists of $r=r_{+}$ and
$r=\infty$. Using the equations of motion $\left(  \ref{eq:NLEqs}\right)  $
and%
\begin{equation}
T_{\mu\nu}=\partial_{\mu}\phi\partial_{\nu}\phi-\frac{1}{2}g_{\mu\nu}\left(
\partial\phi\right)  ^{2}+e^{\alpha\phi^{2}}\left(  F_{\mu\rho}F_{\nu}%
^{\;\rho}-\frac{1}{4}g_{\mu\nu}F_{\rho\sigma}F^{\rho\sigma}\right)  ,
\label{eq:enetgy momentum tensor}%
\end{equation}
we find that the identity $\left(  \ref{eq:Guass's law}\right)  $ is reduced
to
\begin{equation}
M=2TS+Q\Phi-e^{-\delta_{0}}\frac{r_{+}^{3}}{L^{2}}+\int_{r_{+}}^{\infty
}dre^{-\delta\left(  r\right)  }\delta^{\prime}\left(  r\right)  \frac{r^{3}%
}{L^{2}}. \label{eq:Smarr relation}%
\end{equation}
With the help of $P=3/8\pi L^{2}$ and eqn. $\left(  \ref{eq:volume}\right)  $,
the above equation becomes the Smarr relation $\left(  \ref{eq:SR}\right)  $.

\subsection{Free Energy}

The free energy plays a crucial role in studying phase structure and
transitions of black holes. The free energy of a black hole can be obtained by
computing the associated Euclidean path integral, which is related to a
thermal partition function. Specifically, the thermal partition function is
usually computed in the semiclassical approximation by exponentiating the
on-shell Euclidean action $S_{\text{on-shell}}^{E}$,
\begin{equation}
Z\sim e^{-S_{\text{on-shell}}^{E}},
\end{equation}
where $S_{\text{on-shell}}^{E}$ is the on-shell Euclidean action.

For the hairy black hole solutions, $S_{\text{on-shell}}^{E}$ usually diverges
on the AdS boundary, which requires some extra boundary terms to regularize
the bulk action $\left(  \ref{eq:action}\right)  $. The regularized action
$S_{R}$ is given by
\ \cite{Emparan:1999pm,Balasubramanian:1999re,Olea:2005gb,Olea:2006vd,Miskovic:2008ck,Kim:2016dik}
\begin{equation}
S_{R}=S_{\text{bulk}}+S_{\text{GH}}+S_{\text{ct}}\text{,}
\label{eq:Reg. action}%
\end{equation}
where the Gibbon-Hawking boundary term $S_{\text{GH}}$ is introduced to render
the variational principle well-posed, and the counterterm $S_{\text{ct}}$ is
constructed to cancel divergences on asymptotic boundaries. Specifically,
these two boundary terms are evaluated on the hypersurface at the spatial
infinity with the induced metric $\gamma$,
\begin{align}
S_{\text{GH}}  &  =-\frac{1}{8\pi}\int d^{3}x\sqrt{-\gamma}\Theta,\nonumber\\
S_{\text{ct}}  &  =\frac{1}{8\pi}\int d^{3}x\sqrt{-\gamma}\left(  \frac{2}%
{L}+\frac{L}{2}R_{3}\right)  , \label{eq:boundary terms}%
\end{align}
where $\Theta$ is the trace of the extrinsic curvature, and $R_{3}$ is the
scalar curvature of the induced metric $\gamma$. Using the equations of motion
$\left(  \ref{eq:NLEqs}\right)  $ and the boundary conditions $\left(
\ref{eq:BC}\right)  $, we obtain
\begin{align}
S_{\text{bulk, on-shell}}^{E}  &  =\frac{1}{T}\left(  \left.  \frac
{e^{-\delta\left(  r\right)  }r^{2}N^{\prime}\left(  r\right)  -2e^{-\delta
\left(  r\right)  }r^{2}N\left(  r\right)  \delta^{\prime}\left(  r\right)
}{4}\right\vert _{r=+\infty}-TS-Q\Phi\right)  ,\nonumber\\
S_{\text{GH, on-shell}}^{E}  &  =\left.  -\frac{1}{T}\left[  \frac
{e^{-\delta\left(  r\right)  }r^{2}N^{\prime}\left(  r\right)  -2e^{-\delta
\left(  r\right)  }r^{2}\delta^{\prime}\left(  r\right)  N\left(  r\right)
}{4}+e^{-\delta\left(  r\right)  }\left(  r-2M+\frac{r^{3}}{L^{2}}\right)
\right]  \right\vert _{r=+\infty},\nonumber\\
S_{\text{ct, on-shell}}^{E}  &  =\left.  \frac{e^{-\delta\left(  r\right)  }%
}{T}\left(  \frac{r^{3}}{L^{2}}+r-M\right)  \right\vert _{r=+\infty}.
\end{align}
The free energy of hairy black hole solutions is then given by
\begin{equation}
F=TS_{\text{R, on-shell}}^{E}=M-TS-Q\Phi. \label{eq:free energy 1}%
\end{equation}
It is deserved to mention that the variational problem is well-defined only
when the potential $\Phi$ is fixed on the boundaries, which means that the
free energy $\left(  \ref{eq:free energy 1}\right)  $ is properly used in a
grand canonical ensemble at a constant potential. On the other hand, the
electric charge $Q$ is fixed in a canonical ensemble. To construct an
appropriate free energy in a canonical ensemble, the regularized action should
be supplied with an additional boundary term \cite{Wang:2018xdz},
\begin{equation}
S_{\text{surf}}=-\frac{1}{4}\int d^{3}x\sqrt{\gamma}f\left(  \phi\right)
F^{\mu\nu}n_{\mu}A_{\nu}.
\end{equation}
This surface term gives an extra contribution to the on-shell Euclidean
action, $S_{\text{surf, on-shell}}^{E}=\frac{Q\Phi}{T}$, which leads to the
free energy in a canonical ensemble,
\begin{equation}
F=M-TS. \label{eq:free energy 2}%
\end{equation}

\section{Phase Structure and Transitions}

\label{sec:phase structure and transition}

In this section, we investigate phase structure and transitions of RNAdS and
hairy black holes in a grand canonical ensemble and a canonical ensemble. For
later convenience, we define the following reduced quantities,
\begin{equation}
\tilde{T}=TL\text{, }\tilde{F}=F/L\text{, }\tilde{r}_{+}=r_{+}/L\text{,
}\tilde{Q}=Q/L\text{, }\tilde{M}=M/L,
\end{equation}
where the AdS radius $L=\sqrt{3/\left(  8\pi P\right)  }$. The accuracy of our
numerical results is tested using the Smarr relation, which estimates the
numerical error to be around $10^{-6}$. Without loss of generality, we focus
on $\alpha=5$ in the remainder of this section.

\subsection{Grand Canonical Ensemble}

In a grand canonical ensemble, black holes are maintained at a constant
temperature $T$ and a constant potential $\Phi$. To express thermodynamic
quantities in terms of $T$ and $\Phi$, we first need to find the horizon
radius $r_{+}$ as a function of $T$ and $\Phi$. If the function $r_{+}\left(
T,\Phi\right)  $ is multivalued, there is more than one black hole phase,
corresponding to different branches of $r_{+}\left(  T,\Phi\right)  $.

\begin{figure}[ptb]
\begin{centering}
\includegraphics[scale=0.8]{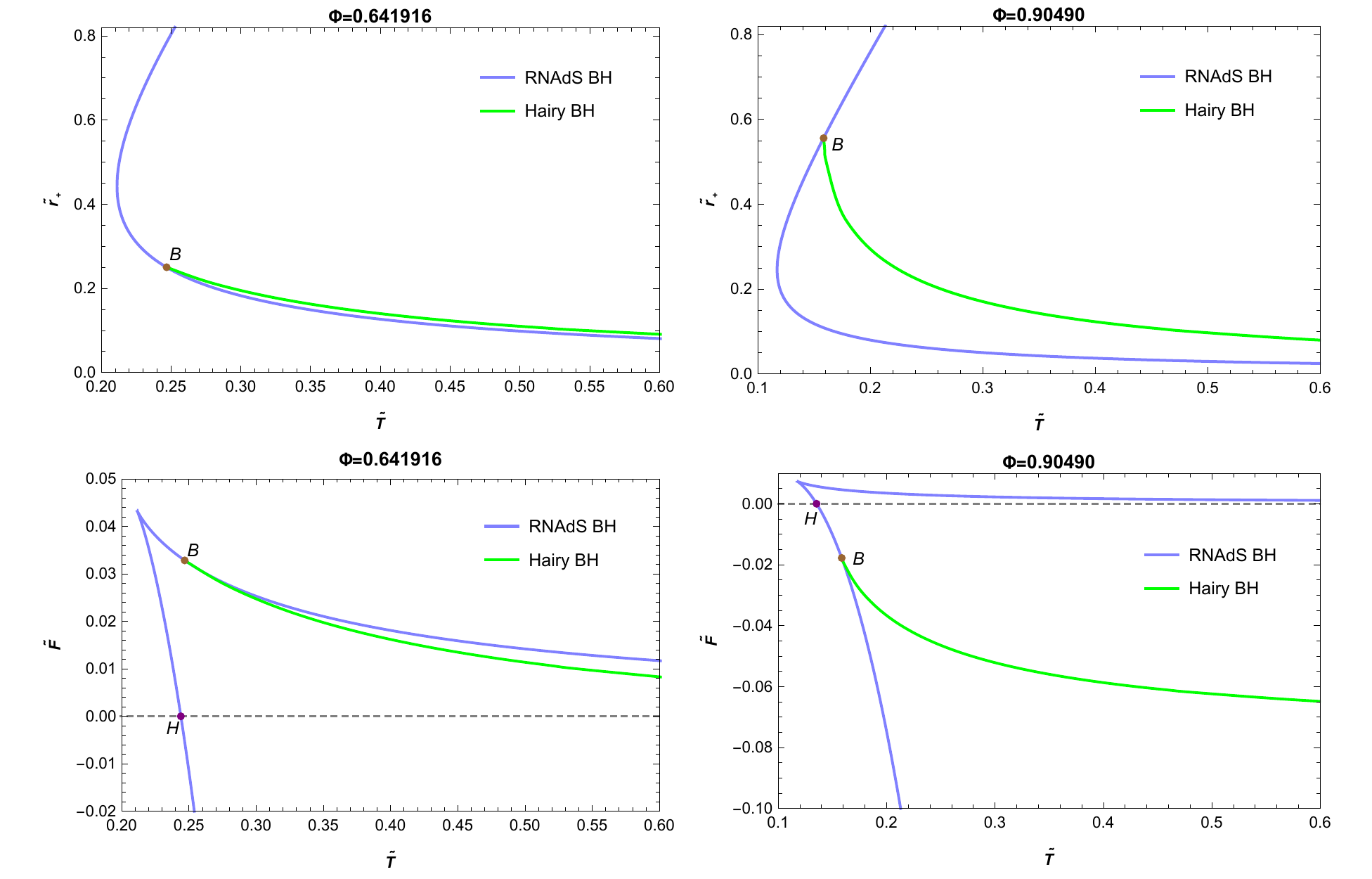}
\par\end{centering}
\caption{Plots of the reduced horizon radius $\tilde{r}_{+}$ (upper row) and
the reduced free energy $\tilde{F}$ (lower row) versus the reduced temperature
$\tilde{T}$ for RNAdS (blue lines) and hairy (green lines) black holes with
$\alpha=5$ in the grand canonical ensemble. Dashed horizontal lines represent
the thermal AdS space, and bifurcation points are labelled by $B$. We consider
two cases with $\Phi=0.641916$ (left column) and $0.90490$ (right column), in
which hairy black holes have only one phase. For RNAdS black holes, $\tilde
{r}_{+}(\tilde{T})$ is double-valued, corresponding to the large RNAdS BH
phase (the branch with a larger horizon radius) and the small RNAdS BH phase
(the branch with a smaller horizon radius). A first-order phase transition
between the thermal AdS space and large RNAdS BH occurs at the point $H$.}%
\label{Fig1}%
\end{figure}

\begin{figure}[ptb]
\includegraphics[scale=0.8]{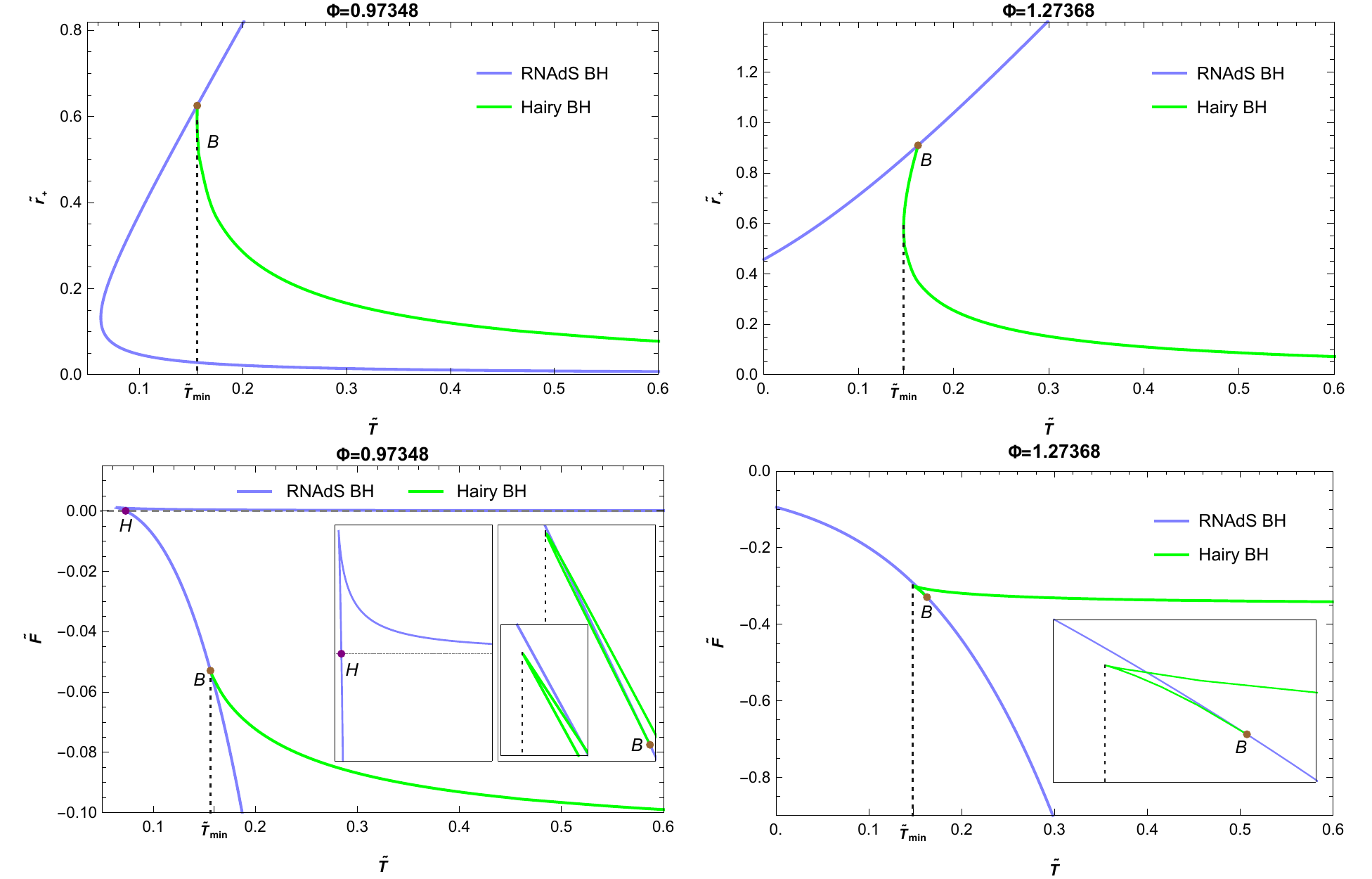}\caption{Plots of the reduced horizon radius
$\tilde{r}_{+}$ (upper row) and the reduced free energy $\tilde{F}$ (lower
row) versus the reduced temperature $\tilde{T}$ for RNAdS (blue lines) and
hairy (green lines) black holes with $\alpha=5$ in the grand canonical
ensemble. Dashed horizontal lines represent the thermal AdS space, and
bifurcation points are labelled by $B$. When $\Phi=0.97348$ and $1.27368$,
there are two phases of hairy black holes, large and small Hairy BHs, which
exist above the minimum temperature $\tilde{T}_{\min}$ (vertical dotted
lines). \textbf{Left column}: $\Phi=0.97348$. RNAdS black holes have two
phases, large and small RNAdS BHs. Increasing the temperature from $\tilde
{T}=0$, the system follows the dashed line of the thermal AdS space until the
point $H$, where a first-order phase transition to large RNAdS BH occurs.
Further increasing $\tilde{T}$, the system jumps to the lower green line of
large Hairy BH, which corresponds to a zeroth-order phase to large Hairy BH.
As $\tilde{T}$ continues to increase, the system follows the lower green line
until it joins the blue line at the bifurcation point $B$, which indicates a
second-order phase transition to large RNAdS BH. In a nutshell, the system
undergoes the Hawking-Page phase transition and a large RNAdS BH/large Hairy
BH/large RNAdS BH reentrant phase transition. \textbf{Right column}:
$\Phi=1.27368$. RNAdS black hole solutions have only one phase, dubbed RNAdS
BH. As $\tilde{T}$ increases, a RNAdS BH/large Hairy BH/RNAdS BH reentrant
phase transition occurs, consisting of a zeroth-order phase transition at
$\tilde{T}_{\text{min}}$ and a second-order phase transition at the
bifurcation point $B$.}%
\label{Fig2}%
\end{figure}

In FIGs. \ref{Fig1} and \ref{Fig2}, we plot the reduced event horizon radius
$\tilde{r}_{+}$ and the reduced free energy $\tilde{F}$ against the reduced
temperature $\tilde{T}$ for RNAdS and hairy black holes with several
representative values of $\Phi$. It is noteworthy that the neutral thermal AdS
space with a fixed potential $\Phi$ is also the solution of eqn. $\left(
\ref{eq:NLEqs}\right)  $, and hence taken into account in these figures. In
the small $\Phi$ regime, FIG. \ref{Fig1} displays $\tilde{r}_{+}(\tilde{T})$
and $\tilde{F}(\tilde{T})$ for RNAdS and hairy black holes with $\Phi
=0.641916$ and $0.90490$. As shown in the upper row, the RNAdS black hole
solutions possess two branches of different horizon radii, dubbed large and
small RNAdS BHs, respectively, whereas the hairy black hole solutions have
only one branch. Moreover, the hairy black holes bifurcate from the RNAdS
black holes at bifurcation points $B$, which determine the minimum temperature
of the hairy black holes. The lower row shows that the hairy black hole phase
cannot be the globally stable against large RNAdS BH (the branch with a larger
horizon radius) since large RNAdS BH always has a lower free energy than the
hairy black hole. For a temperature greater/less than that of the point $H$,
large RNAdS BH/thermal AdS space is globally stable, which leads to the
first-order Hawking-Page phase transition at the point $H$.

For a large enough value of $\Phi$, hairy black holes can possess two phases,
namely the large and small Hairy BH phases. In FIG. \ref{Fig2}, we present
$\tilde{r}_{+}$ and $\tilde{F}$ as functions of $\tilde{T}$ for RNAdS and
hairy black holes with $\Phi=0.97348$ and $1.27368$. It is observed that there
are two phases for the hairy black holes in both cases, while the RNAdS black
holes have one (two) phase(s) for $\Phi=0.97348\left(  1.27368\right)  $.
Moreover, the hairy black holes have a minimum temperature $\tilde{T}_{\min}$,
and large Hairy BH coexists with small Hairy BH between $\tilde{T}=\tilde
{T}_{\min}$\ and the bifurcation points $B$, where large Hairy BH and RNAdS
black holes merge. When $\Phi=0.97348$, RNAdS black hole solutions have two
branches, corresponding to large and small RNAdS BHs. As $\tilde{T}$ increases
from zero, there occurs a first-order phase transition from the thermal AdS
space to large RNAdS BH at the purple point $H$ . Subsequently, the left inset
of the lower panel shows that the globally stable phase jumps to large Hairy
BH from large RNAdS BH, which signals a zeroth-order phase transition
occurring at $\tilde{T}=\tilde{T}_{\text{min}}$. Further increasing $\tilde
{T}$, we observe that the globally stable phase remains large Hairy BH until
the bifurcation point $B$, where the system undergoes a second-order phase
transition and returns to large RNAdS BH. Note that hairy and RNAdS black
holes have the same entropy at the bifurcation point, thus indicating the
phase transition at the bifurcation point is second-order. In short, a RNAdS
BH/hairy BH/RNAdS BH reentrant phase transition is observed as $\tilde{T}$
increases. When $\Phi=1.27368$, RNAdS black hole solutions have only one
phase, whose free energy is always smaller than that of the thermal AdS space.
So there is no Hawking-Page first-order phase transition, However, as shown in
the inset of the lower panel, a RNAdS BH/large Hairy BH/RNAdS BH reentrant
phase transition still occurs.

In addition, it is of interest to consider local thermal stability of black
hole solutions against thermodynamic fluctuations. In a grand canonical
ensemble, a specific heat at constant potential and pressure,
\begin{equation}
C_{\Phi,P}=T\left(  \frac{\partial S}{\partial T}\right)  _{\Phi,P}%
=\frac{3\tilde{r}_{+}\tilde{T}}{4P}\left(  \frac{\partial\tilde{r}_{+}%
}{\partial\tilde{T}}\right)  _{\Phi,P},
\end{equation}
can be used to analyze the thermal stability. Specially, the thermal stability
of a black hole phase follows from $C_{\Phi,P}>0$ (or equivalently,
$\partial\tilde{r}_{+}/\partial\tilde{T}>0$). From the upper rows of FIG.
\ref{Fig1} and FIG. \ref{Fig2}, we notice that the globally stable phases of
RNAdS and hairy black holes possess a positive $C_{Q}$, and are thermally stable.

\begin{figure}[ptb]
\includegraphics{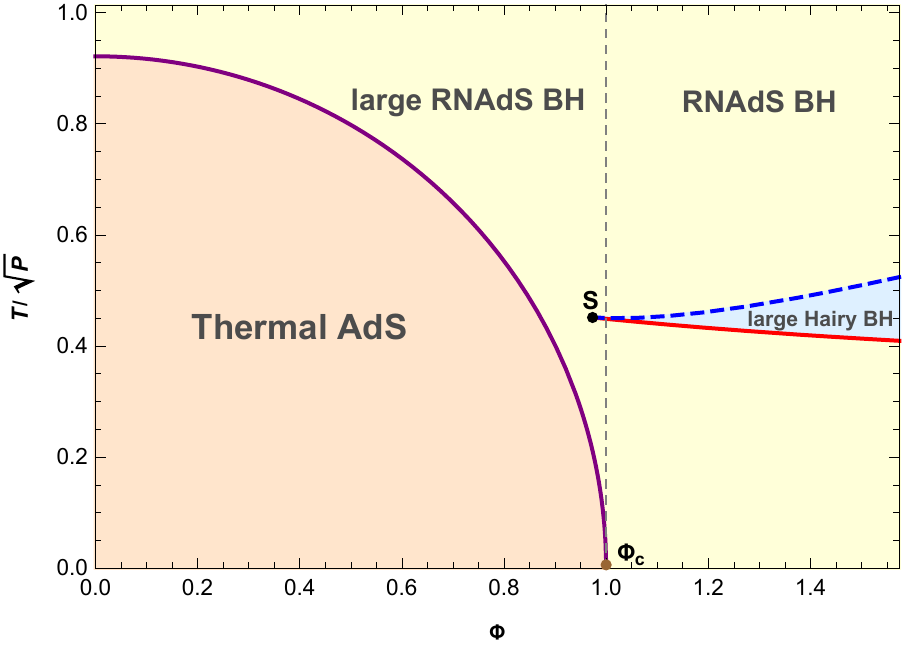}\caption{Phase diagram of the grand canonical ensemble
of hairy and RNAdS black holes with $\alpha=5$ in the $\Phi$-$T/\sqrt{P}$
plane. A first-order phase transition line (purple line) separates the thermal
AdS space and large RNAdS BH, and terminates at $\Phi=\Phi_{\text{c}}$. In the
large $\Phi$ regime, large Hairy BH appears, and is bounded by a zeroth-order
phase transition line (red line) and a second-order one (blue dashed line).}%
\label{Fig3}%
\end{figure}

\begin{figure}[ptb]
\includegraphics[scale=0.85]{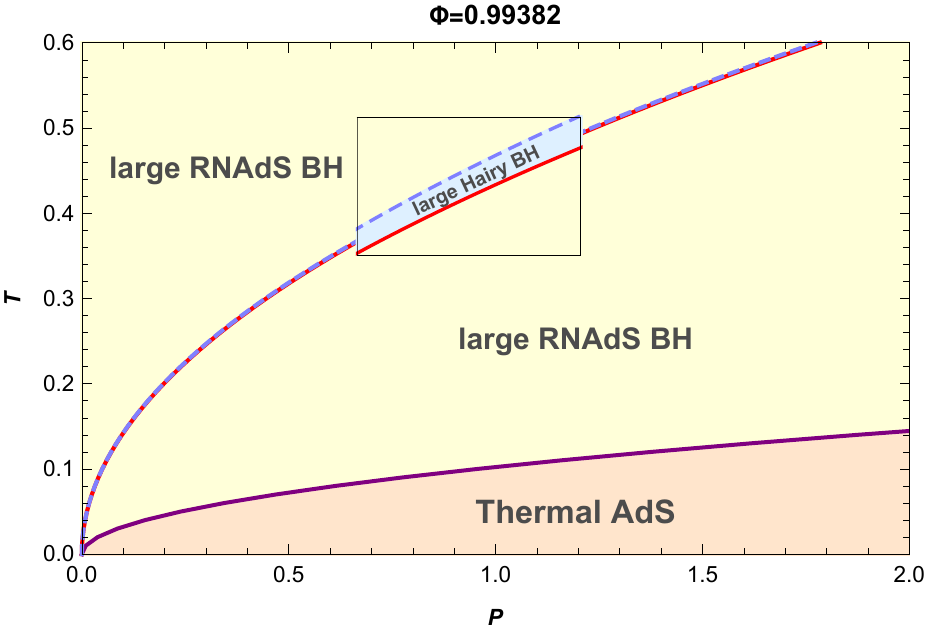}
\includegraphics[scale=0.85]{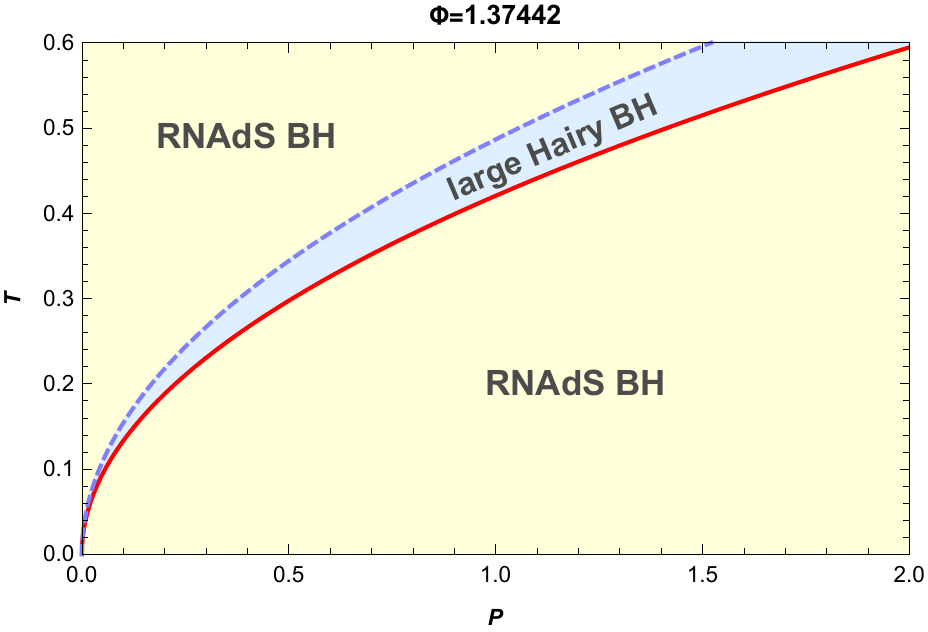}\caption{Phase diagrams of the grand
canonical ensemble of hairy and RNAdS black holes with $\alpha=5$ in the
$P\text{-}T$ plane.\textbf{ Left panel}: $\Phi=0.99382$. There is a
first-order phase transition line (purple line) between the thermal AdS space
and large RNAdS BH, which is semi-infinite in the $P\text{-}T$ plane and
reminiscent of the solid/liquid phase transition. The inset exhibits the large
Hairy BH phase between the zeroth-order (red line) and second-order (blue
dashed line) phase transition lines. \textbf{Right panel}: $\Phi=1.37442$. The
thermal AdS space and the first-order Hawking-Page phase transition are
absent. RNAdS black holes have only one phase, namely RNAdS BH. The blue
strip, corresponding to large Hairy BH, resembles that in the left panel, but
with a larger width.}%
\label{Fig4}%
\end{figure}

To better illustrate phase structure and transitions of hairy and RNAdS black
holes, we plot phase diagrams in the $\Phi\text{-}T/\sqrt{P}$ and $P\text{-}T$
planes in FIG. \ref{Fig3} and FIG. \ref{Fig4}, respectively, where $\alpha=5$.
These phase diagrams exhibit the globally stable phases, which possess the
lowest free energy, and the phase transitions between them. In FIG. \ref{Fig3}
and FIG. \ref{Fig4}, purple lines correspond to first-order phase transitions
between the thermal AdS space and large RNAdS BH, red lines to zeroth-order
phase transitions between RNAdS black holes and large Hairy BH, and blue
dashed lines to second-order phase transitions between large Hairy BH and
RNAdS black holes.

In the phase diagram in the $\Phi\text{-}T/\sqrt{P}$ plane, FIG. \ref{Fig3}
shows that the first-order phase transition line separating the thermal AdS
phase and large RNAdS BH exists when $\Phi<\Phi_{\text{c}}=1$. It is
noteworthy that there are two branches of RNAdS black hole solutions for
$\Phi<\Phi_{\text{c}}$, and only one branch for $\Phi>\Phi_{\text{c}}$. In the
large $\Phi$ regime, the large Hairy BH phase can be globally stable for some
range of $\tilde{T}$, and is bounded by zeroth-order and second-order phase
transition lines, which merge at the point $S$.

We depict phase diagrams in the $P\text{-}T$ plane for $\Phi=0.99382$ and
$\Phi=1.37442$ in FIG. \ref{Fig4}. When $\Phi=0.99382$, the left panel of FIG.
\ref{Fig4} displays that the first-order phase transition line is
semi-infinite in the $P\text{-}T$ plane, which is reminiscent of the
solid/liquid phase transition. The zoomed-in inset shows that large Hairy BH
exists for a narrow range of $\tilde{T}$, and is separated from RNAdS black
holes by the zeroth-order and second-order phase transition lines. On the
other hand, when $\Phi=1.37442$, the thermal AdS phase is never globally
preferred, and RNAdS black hole solutions are always single-valued. So in the
right panel of FIG. \ref{Fig4}, only RNAdS BH and large Hairy BH, as well as
the associated zeroth-order and second-order phase transitions, are presented.

\subsection{Canonical Ensemble}

In a canonical ensemble with\ fixed black hole charge $Q$ and temperature $T$,
we use the free energy $\left(  \ref{eq:free energy 2}\right)  $ to study
phase structure and transitions of RNAdS and hairy black holes. Furthermore, a
specific heat at constant charge and pressure,%
\begin{equation}
C_{Q,P}=T\left(  \frac{\partial S}{\partial T}\right)  _{Q,P}=\frac{3\tilde
{r}_{+}\tilde{T}}{4P}\left(  \frac{\partial\tilde{r}_{+}}{\partial\tilde{T}%
}\right)  _{Q,P},
\end{equation}
is used to investigate the local thermal stability of black holes in the
canonical ensemble.

\begin{figure}[ptb]
\includegraphics[scale=0.6]{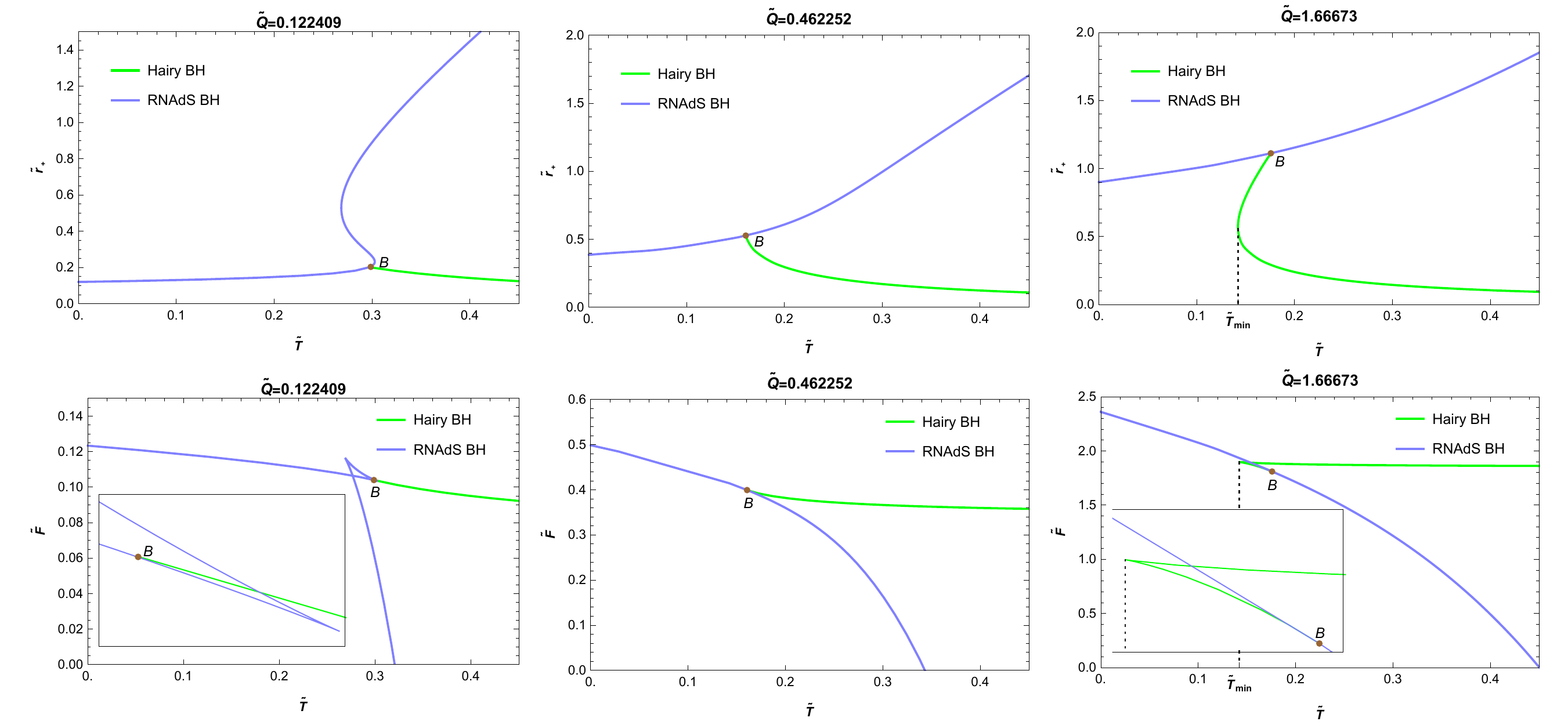}\caption{Plots of the reduced horizon
radius $\tilde{r}_{+}$ (upper row) and the reduced free energy $\tilde{F}$
(lower row) against the reduced temperature $\tilde{T}$ for RNAdS (blue lines)
and hairy (green lines) black holes with three values of reduced charge
$\tilde{Q}$ in the canonical ensemble. Bifurcation points are marked by $B$.
\textbf{Left column}: $\tilde{Q}=0.122409$. Three branches of RNAdS black hole
solutions coexist in some range of $\tilde{T}$, where a first-order phase
transition between large RNAdS BH and small RNAdS BH occurs. Hairy black holes
bifurcate from RNAdS black holes at the point $B$ with a higher free energy,
and therefore are not the globally preferred phase. \textbf{Center column}:
$\tilde{Q}=0.462252$. There is only one branch for RNAdS and hairy black hole
solutions, and no phase transition occurs. \textbf{Right column}: $\tilde
{Q}=1.66673$. RNAdS black holes coexist with two branches of hairy black holes
in a certain range of $\tilde{T}$, where large Hairy BH is globally preferred.
A RNAdS BH/large Hairy BH/RNAdS BH reentrant phase transition, consisting of
zeroth-order and second-order phase transitions, occurs as $\tilde{T}$
increases.}%
\label{Fig5}%
\end{figure}

We plot the reduced horizon radius $\tilde{r}_{+}$ and the free energy
$\tilde{F}$ as functions of reduced temperature $\tilde{T}$ for RNAdS and
hairy black holes with three representative values of $\tilde{Q}$ in Fig.
\ref{Fig5}, where we have $\alpha=5$. The presence of multivalued functions
$\tilde{r}_{+}(\tilde{T})$ indicates that black hole solutions can possess
multi branches of different horizon radii, which may lead to phase
transitions. The lower row exhibits the free energy $\tilde{F}$ against the
temperature $\tilde{T}$, which shows globally preferred phases. In the left
column, there are three branches of RNAdS black holes in some range of
$\tilde{T}$, dubbed large, intermediate and small RNAdS BHs depending on their
sizes of horizon radius. Whereas hairy black holes, emerging from the
bifurcation point $B$, have only one branch of solutions. It shows that a
first-order phase transition occurs between large and small RNAdS BH phases,
both of which have positive $C_{Q,P}$, and hence are thermally stable. In the
middle column of FIG. \ref{Fig5}, RNAdS and hairy black holes both have a
single phase. Moreover, the RNAdS black holes are always globally preferred
over the hairy black holes, and therefore there is no phase transitions. In
the right column of FIG. \ref{Fig5}, there are only one branch of RNAdS black
hole solutions and two branches of hairy black hole solutions. The inset
displays that, as $\tilde{T}$ increases from $0$, the system undergoes a
zeroth-order phase transition from RNAdS BH to large Hairy BH at $\tilde
{T}=\tilde{T}_{\min}$ and a second-order phase transition from large Hairy BH
to RNAdS BH at the bifurcation point $B$, corresponding to a RNAdS BH/large
Hairy BH/RNAdS BH reentrant phase transition. Note that both globally
preferred phases are thermally stable with a positive $C_{Q,P}$.

\begin{figure}[ptb]
\includegraphics[scale=0.85]{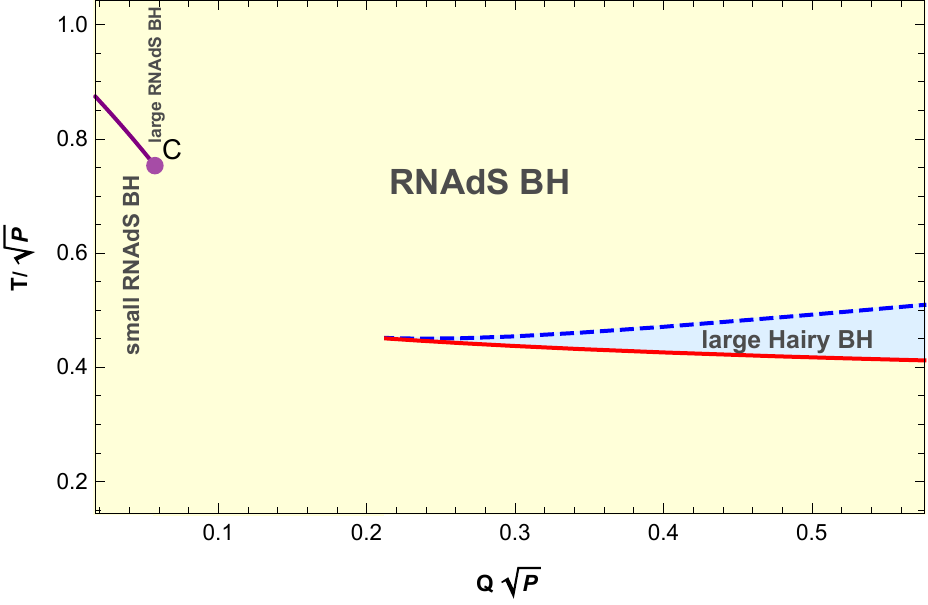}
\includegraphics[scale=0.85]{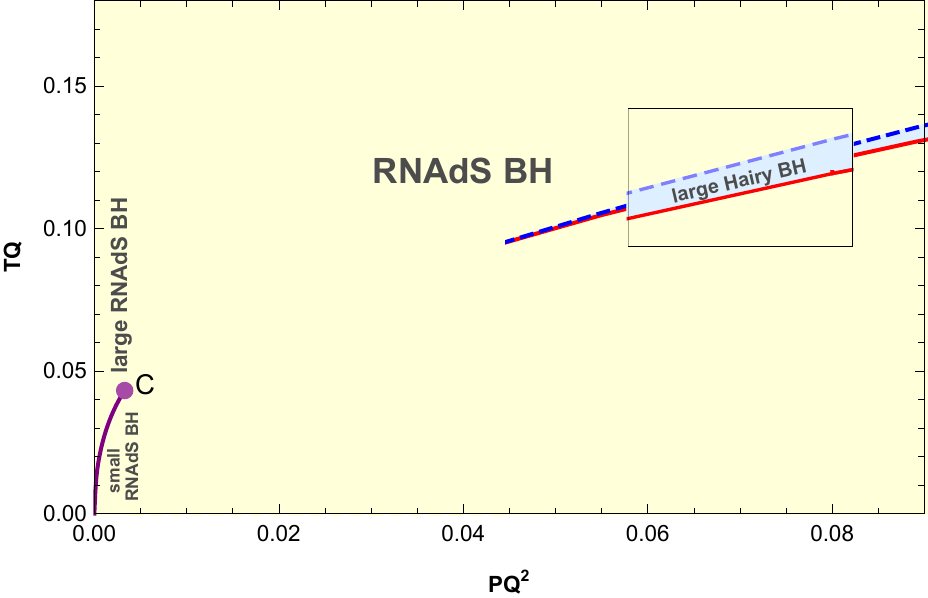}\caption{Phase diagrams of the
canonical ensemble of RNAdS and hairy black holes with $\alpha=5$ in the
$Q\sqrt{P}\text{-}T/\sqrt{P}$ (left panel) and $PQ^{2}\text{-}TQ$ (right
panel) planes. There is a first-order phase transition (purple line) between
small and large RNAdS BHs, which terminates at a critical point $C$. Past the
critical point, RNAdS black holes have only one phase, and large Hairy BH can
be the globally stable phase. For a large value of $Q\sqrt{P}$ or $PQ^{2}$,
the system undergoes a RNAdS BH/large Hairy BH/RNAdS reentrant phase
transition, which is composed of zeroth-order (red line) and second-order
(blue dashed line) phase transitions, with the temperature increasing.}%
\label{Fig6}%
\end{figure}

In FIG. \ref{Fig6}, the phase diagrams display the globally stable phases with
the lowest free energy in the $Q\sqrt{P}\text{-}T/\sqrt{P}$ and $PQ^{2}%
\text{-}TQ$ planes. For a small value of $Q\sqrt{P}$ or $PQ^{2}$, a
first-order phase transition separates small and large RNAdS BHs and
terminates at the critical point, which is reminiscent of the liquid/gas phase
transition. However, as $Q\sqrt{P}$ or $PQ^{2}$ increases, the large Hairy BH
phase can be the globally preferred state. In particular, large Hairy BH is
bounded by second-order and zeroth-order phase transition lines, which
corresponds to a RNAdS BH/large Hairy BH/RNAdS BH reentrant phase transition.

\section{Conclusions}

\label{sec:Discussion-and-Conclusions}

In this paper, we first briefly introduced asymptotically AdS hairy black
holes in the EMS model \cite{Guo:2021zed}, where a non-minimal coupling
function $f\left(  \phi\right)  =e^{\alpha\phi^{2}}$ between the scalar and
Maxwell fields was considered. We then derived the first law of thermodynamics
following the covariant construction associated with the diffeomorphism
symmetry, and obtained the Smarr relation by computing a Komar integral with
respect to a time-like Killing vector. Focusing on the extended phase space,
we found that the conjugate thermodynamic volume of hairy black holes is
different from the \textquotedblleft naive\textquotedblright\ geometric volume
(i.e., $4\pi r_{+}^{3}/3$), and computed the regularized on-shell Euclidean
action to obtain the free energy of black holes in canonical and grand
canonical ensembles.

In the EMS model, there exists some parameter region where hairy and RNAdS
black holes coexist, which leads to complicated phase structure in canonical
and grand canonical ensembles. In the grand canonical ensemble, a thermal
AdS/large RNAdS BH first-order phase transition occurs for a small $\Phi$,
which composes a semi-infinite coexistence line in the $P\text{-}T$ diagram
and is reminiscent of the solid/liquid phase transition. In the canonical
ensemble, there is a small RNAdS BH/large RNAdS BH first-order phase
transition, which terminates at a critical point and hence resembles the
liquid/gas phase transition. More interestingly, RNAdS BH/hairy BH/RNAdS BH
reentrant phase transitions, which consist of zeroth-order and second-order
phase transitions, were observed in both ensembles. As discussed in the
Introduction, the reentrant phase transitions that have been reported in the
context of black holes usually include zeroth-order and first-order phase
transitions. However, the reentrant phase transitions in our paper were found
to consist of zeroth-order and second-order phase transitions. The
second-order phase transition between RNAdS and hairy black holes may provide
important holographic applications.

\begin{acknowledgments}
We are grateful to Qingyu Gan and Feiyu Yao for useful discussions and
valuable comments. This work is supported in part by NSFC (Grant No. 11875196,
11375121, 11947225 and 11005016).
\end{acknowledgments}

\bibliographystyle{unsrturl}
\bibliography{ref}

\end{document}